\documentclass[%
reprint,
superscriptaddress,
 amsmath,amssymb,
aps,
 prb,
]{revtex4-2}

\usepackage{hyperref}
\usepackage[english]{babel} 
\usepackage{amssymb}
\usepackage{amsmath}
\usepackage{txfonts}
\usepackage{mathdots}
\usepackage[classicReIm]{kpfonts}
\usepackage{graphicx}
\usepackage[dvipsnames]{xcolor}
\def\arrvline{\hfil\kern\arraycolsep\vline\kern-\arraycolsep\hfilneg}
\usepackage{dcolumn}
\usepackage{bm}

\begin{document}
\preprint{APS/123-QED}
\title{Nuclear spin-spin interactions in CdTe probed by zero and ultra-low-field optically detected NMR.}

\author{V.~M.~Litvyak}
\affiliation{Spin Optics Laboratory, St. Petersburg State University, 198504 St. Petersburg, Russia}
\author{P.~Bazhin}
\affiliation{Spin Optics Laboratory, St. Petersburg State University, 198504 St. Petersburg, Russia}
\author{R.~André}
    \affiliation{Universit\'e Grenoble Alpes, CNRS, Institut N{\'{e}}el, 38000 Grenoble, France}
\author{M.~Vladimirova}
\affiliation{Laboratoire Charles Coulomb, UMR 5221 CNRS/Universit\'{e} de Montpellier, France}
\author{K.~V.~Kavokin}
\affiliation{Spin Optics Laboratory, St. Petersburg State University, 198504 St. Petersburg, Russia}

\begin{abstract}
Nuclear magnetic resonance (NMR) is particularly relevant for studies of internuclear spin coupling at zero and ultra-low fields (ZULF), where spin-spin interactions dominate over Zeeman ones. 
Here we report on ZULF NMR  in CdTe. 
In this semiconductor all magnetic isotopes have spin $I = 1/2$, so that internuclear interactions are never overshadowed by quadrupole effects.
Our experiments rely on warm-up spectroscopy, a technique that combines optical pumping, additional cooling via adiabatic demagnetisation, and detection of the oscillating magnetic field-induced warm-up of the nuclear spin system via Hanle effect. 
 We show that NMR spectra exhibit a rich fine structure, consistent with the  low abundance of magnetic isotopes in CdTe, their zero quadrupole moments, as well as direct and indirect interactions between them. 
 A model assuming that the electromagnetic radiation is absorbed by {nuclear spin} clusters composed of up to $5$ magnetic isotopes  allows us to reproduce the shape of a major part of the measured spectra.   
\end{abstract}

\date{\today}
\maketitle
\section{Introduction }

Nuclear spin interactions provide a wealth of information about the connectivity and spatial ordering of atoms in solid materials \cite{Abragam}, they determine thermodynamics of the nuclear spin system \cite{Goldman} and are of crucial importance in many domains, spreading from physics, chemistry and biology \cite{kemp_nmr_1986} to quantum information processing \cite{oliveira_5_2007}. 
Experimentally these interactions are probed via nuclear magnetic resonance (NMR) experiments, where absorption of the electromagnetic radiation as a function of its {frequency} is measured at relatively high static magnetic field.

In semiconductors, nuclei with non-zero magnetic moments participate in direct dipole-dipole and indirect (exchange and pseudodipolar) interactions \cite{Paget1977}. 
The strength of these internuclear interactions is deduced from { the shape} of NMR lines at high 
external magnetic fields, where nuclear Zeeman energies  are many orders of magnitude higher than the spin-spin ones.

The magnetic dipole-dipole interaction is the most long-range type of internuclear spin-spin interactions. Even at high magnetic field it is non-trivial to determine theoretically the precise shape of the absorption spectrum due to dipole-dipole coupling, but its contribution to the widths of NMR lines  was calculated by Van Vleck in 1948 \cite{van_vleck_dipolar_1948}.

Indirect interactions (exchange and pseudodipolar) were included into consideration somewhat later. These interactions are mediated by valence electrons, and  differ from zero only for the nearest neighbor nuclei. 
For example, the indirect exchange interaction was experimentally detected from the broadening of NMR lines in GaSb \cite{shulman_nuclear_1955} and in GaAs \cite{shulman_nuclear_1958}. The theoretical description was provided in  \cite{anderson_considerations_1955}. The pseudodipolar interaction was first mentioned in Ref. \onlinecite{bloembergen_nuclear_1955}. 
Then it was found that the pseudodipolar interaction in GaAs crystals cancels a considerable part of the dipole-dipole contribution to the second moment of NMR lines  \cite{hester_nuclear-magnetic-resonance_1974,Litvyak_local_2023}. 
The values of the isotope parameters and interaction constants in GaAs are reported in Tables~\ref{tab:tab1},\ref{tab:tab2}. Here $D$ characterises direct dipole-dipole coupling between nearest neighbors, $J_{\perp}$ and $J_{\parallel}$ account for the anisotropy of the exchange interaction between two neighboring nuclei with respect  to the axis of the interatomic bonds, see Sec.~\ref{sec:model}.

Because in traditional high-field NMR  spin-spin interaction energies are many orders of magnitude smaller than Zeeman ones,
the only terms of the spin-coupling Hamiltonians that may be observed in such experiments are those that commute with the Zeeman Hamiltonian.
Furthermore, spin-spin interactions usually make more subtle contributions to the zero-field NMR spectrum than quadrupole interactions. Therefore spin-spin interactions can be difficult to determine in the systems with spins greater than $1/2$ if any non-intentional residual electric field gradients are present \cite{Litvyak_warm-up_2021,Litvyak_local_2023}. 
However, because NMR sensitivity depends on magnetization magnitude and precession frequency, and  both  decrease {when the magnetic field is decreased}, measurements of the spin-spin interactions at low fields are particularly challenging.

\begin{figure}[!h]
	\includegraphics[width=3.4in]{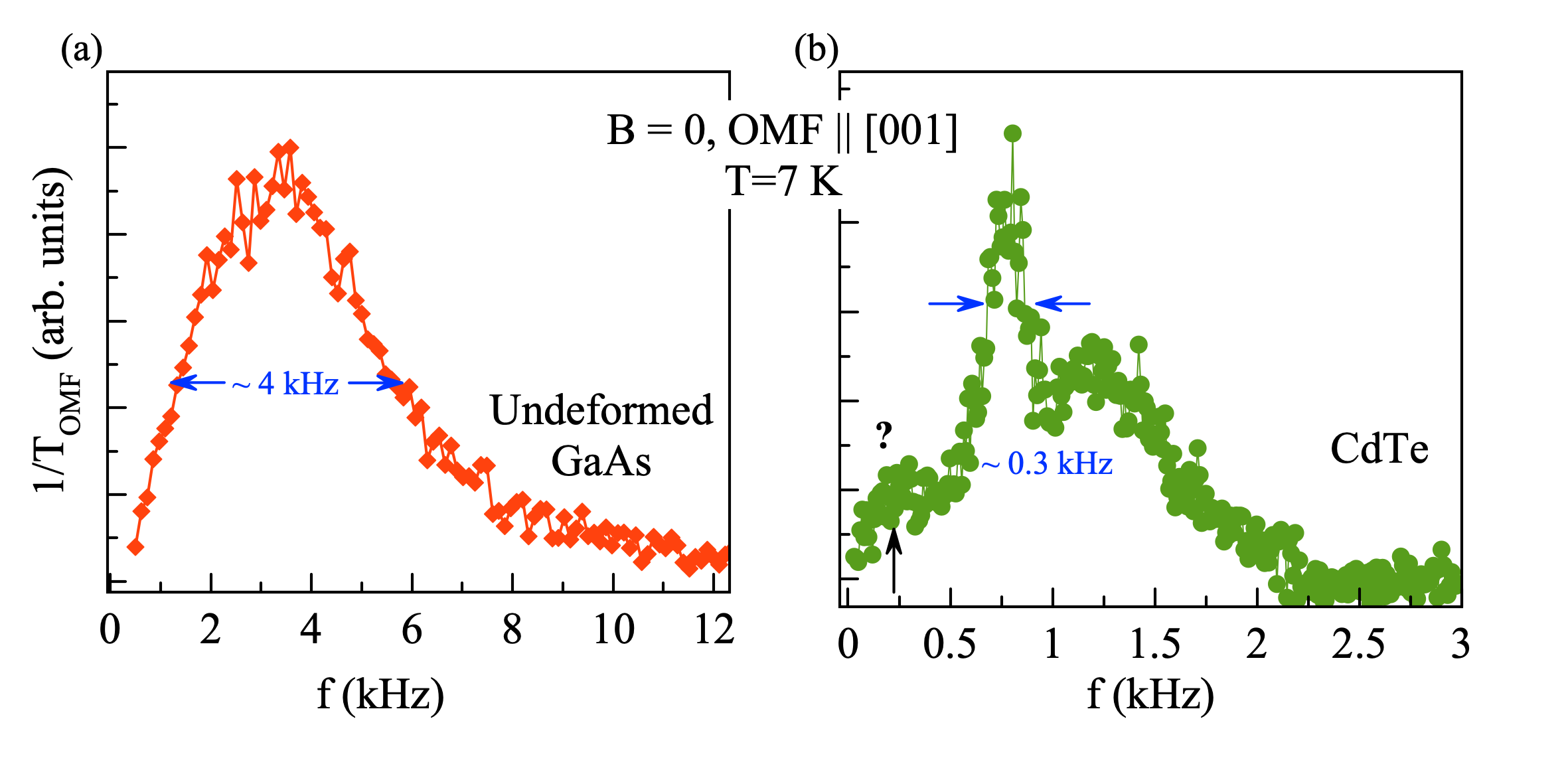}
	\caption{ Absorption {spectra} of optically cooled NSS in zero magnetic field in bulk n-GaAs, see  Ref. \onlinecite{Litvyak_local_2023} (a) and  in wide CdTe quantum well studied in this work (b). 
}
	\label{fig:fig1}
\end{figure}

Starting with the pioneering works in 1980s \cite{zax_heteronuclear_1984,Kalevich83}, a trend has developed toward optically detected NMR (ODNMR) using submicrotesla fields or even no external field at all.
In semiconductors, where deep cooling of the nuclear spin system can be reached by optical pumping followed by adiabatic demagnetisation to zero  field, a specific approach to zero and ultra-low field (ZULF) NMR, termed as warm-up spectroscopy, has been proposed in Ref. \cite{Kalevich83}.  The warm-up spectroscopy has been further developed in \cite{Litvyak_warm-up_2021}. The method is based on the measurements of the oscillating magnetic field (OMF)-induced modifications of the nuclear spin temperature.   This concept is particularly relevant in ZULF regime, where it has been successfully employed for the description of a plethora of the experimental data \cite{AbragamProctor,Vladimirova2018,Litvyak_warm-up_2021,Litvyak_local_2023}. However,  only few ZULF ODNMR experiments in semiconductors are currently available, and they mainly address  GaAs based samples \cite{giri_nondestructive_2013,Litvyak_warm-up_2021,vladimirova_simultaneous_2022,Litvyak_local_2023}.

The absorption spectrum of the lightly n-doped GaAs crystal measured in Ref.~\onlinecite{Litvyak_local_2023}  by warm-up technique at zero magnetic field is shown in Fig.~\ref{fig:fig1}~(a). It presents a single  broad peak at the frequency $f \approx 3.5$~kHz, with full width at half maximum (FWHM) $\approx 4$~kHz. Unfortunately, theoretical  calculation of such {a} spectrum in ZULF regime and thus identification of different kinds of the spin-spin interactions on the basis of comparison with the data is prohibitively complex, due to long-range nature of the dipole-dipole coupling and potentially nonzero quadrupole contribution.
%

%



In this context, CdTe is a very promising material. 
It has a zinc-blend crystal structure, similar to GaAs, but its nuclear spin system (NSS) is very dilute, only about one-third of the nuclei {have} a non-zero magnetic moment. 
The direct dipole-dipole coupling is  rather weak, as compared to GaAs, where all the isotopes are magnetic \cite{van_vleck_dipolar_1948}. On the other hand, high atomic numbers  favor  stronger indirect spin-spin coupling, see Table \ref{tab:tab2}. 
Finally, in contrast with GaAs, all magnetic isotopes have  spin $I=1/2$, so that quadrupole effects {can not} mask internuclear coupling.
The main characteristics of CdTe isotopes with nonzero magnetic moments, as well as known from the literature parameters characterizing their interactions  are summarized in Tables \ref{tab:tab1},  \ref{tab:tab2} \footnote{We neglect in this work $^{123}$Te, a magnetic isotope with very low abundance \cite{harris_nmr_2002}.}. For comparison, the same parameters for GaAs are reported. 

The peculiarities of  nuclear spin-spin interactions in CdTe crystals were discovered  by Nolle and co-workers using traditional high-field NMR spectroscopy ($B=2.114$~T) \cite{koch_77se_1978,nolle_direct_1979,balz_125te_1980}.  These measurements required significant acquisition times,  about $12$~h/spectrum, and provided remarkable results.
Spectra around resonance frequency of  $^{125}$Te for different orientations of the crystal with respect to the external magnetic field appeared to have complex structure consisting of the main line and several satellites, depending on the field orientation. These fine structure was interpreted as being due to spin-spin interactions. The widths of the individual lines appeared to be as small as $\approx 0.2$~kHz, roughly $10$ times smaller than in GaAs crystals. This points out the significance of the difference between CdTe and GaAs  NSS properties.  
From the measured spectra of $^{125}$Te Nolle calculated the constants of  pseudodipolar and exchange interactions reported in  Table \ref{tab:tab2}, but, to the best of our knowledge, no experimental studies of the nuclear spin coupling at zero and low magnetic fields has been reported yet.

 \begin{table}
\begin{tabular}{|>{\centering\arraybackslash}p{1.2cm}|>{\centering\arraybackslash}p{1.2cm}|>{\centering\arraybackslash}p{1.2cm}|>{\centering\arraybackslash}p{1.2cm}|>{\centering\arraybackslash}p{1.2cm}|>{\centering\arraybackslash}p{1.2cm}|}
\hline
 { } & $I$ & $\gamma$ (kHz/G) &   A$_{\mathrm{hf}}$ ($\mathrm{\mu}$eV)  &  $A$ & $P_m$  \\ 
\hline 
${}^{69}$Ga &$3/2$ &  1.03&  43.1  &  0.6 &   {1.0} \\ 
\hline 
${}^{71}$Ga &$3/2$ & 1.29 &  54.8 &     0.4 &   {1.0} \\
\hline 
 ${}^{75}$As &$3/2$ &  0.73 & 43.5 &      {1.0} &   {1.0} \\ 
 \hline 
GaAs      &$3/2$   &  0.93   &   91 &     {2.0} &   {} \\ 
\hline 
\hline 
 $^{111}$Cd &$1/2$ & {-0.91}  & -37.4 &   {0.13} &   {0.91} \\ 
\hline
$^{113}$Cd &$1/2$& -0.95   &   -39.1 &   {0.12} &   {0.91} \\ 
\hline 
$^{125}$Te &$1/2$ & -1.35 &   -45 &    {0.08} & 0.75 \\ 
\hline 
$^{123}$Te &$1/2$ & -1.12 &   -45 &    {0.01} &   {0.75} \\ 
\hline 
CdTe    &$1/2$ & -0.17 &   -13 &     {0.34} &   {}\\ 
\hline 
\end{tabular}
\caption{Magnetic isotope parameters. Gyromagnetic ratio  $\gamma$, hyperfine constant A$_{\mathrm{hf}}$, abundance  $A$, and the probability to have a magnetic neighbor $P_m$.  Average values of $\gamma$ and  A$_{\mathrm{hf}}$, as well as the total abundances of magnetic isotopes per unit cell consisting of two atoms in CdTe and in GaAs are given\cite{harris_nmr_2002}. We neglect in this work $^{123}$Te, a magnetic isotope with very low abundance.}
\label{tab:tab1}
\end{table}

 \begin{table}
\begin{tabular}{|>{\centering\arraybackslash}p{0.22\linewidth}|>{\centering\arraybackslash}p{0.22\linewidth}|>{\centering\arraybackslash}p{0.22\linewidth}|>{\centering\arraybackslash}p{0.22\linewidth}|}
\hline
 { } & $J_{\perp}$ (kHz)  &  $J_{\parallel}$ (kHz) &   D (kHz)  \\ 
\hline 
${}^{69}$Ga/${}^{75}$As &  0.034$^a$, -0.339$^b$ & 0.643$^a$, 0.678$^b$ &  0.339 \\ 
\hline 
${}^{71}$Ga/${}^{75}$As & 0.043$^a$, -0.43$^b$ & 0.817$^a$, 0.86$^b$ &  0.43 \\ 
\hline 
 $^{111}$Cd/$^{125}$Te   &   0.723 &    0.42 & 0.364 \\ 
 \hline 
 $^{111}$Cd/$^{125}$Te  & 0.765 &   {0.435} &   0.38 \\ 
\hline 
\end{tabular}
\caption{Spin-spin coupling constants from Ref. \onlinecite{nolle_direct_1979} in CdTe and from \cite{cueman_pseudodipolar_1976}$^a$, \cite{Litvyak_local_2023}$^b$ in GaAs.  }
\label{tab:tab2}
\end{table}

In this work we report on ZULF ODNMR experiments in CdTe.  Fig.~\ref{fig:fig1}~(b) shows the absorption spectrum measured at zero magnetic field.  As for GaAs (Fig.~\ref{fig:fig1}~(a)), the experiments are performed using warm-up spectroscopy \cite{Litvyak_warm-up_2021}, but in CdTe we observe multiple absorption lines, with much smaller FWHM, $\approx 0.3$~kHz, consistent with dipole-dipole interaction between nearest neighbors.

In the absence of quadrupole interactions, the observed fine structure must be related to  the internuclear spin coupling.  We propose a model based on the hypothesis that NSS comprises mainly isolated   {noninteracting} spins and small clusters consisting of up to $5$ magnetic isotopes. The long-range interaction between clusters is neglected. Under these assumptions one can reduce the problem of the interactions among all nuclei  to the problem of the ensemble of interactions among a small number of nearest neighbors inside so-called nuclear spin clusters. 

Using the hypothesis presented above, we calculate the spectral dependence of the NSS warm-up rate induced by the OMF  at zero and low static magnetic fields for different orientations of the OMF with respect to the static field and the crystal axes, and compare these calculations with the experimentally measured spectra. An agreement between the model and the data is achieved in the absence of fitting parameters in most of the studied experimental configurations, suggesting the relevance and validity of the cluster model.

Most of the observed spectral features could be identified within the model on the basis of the internuclear interactions constants deduced from high-field NMR, except from the lowest frequency peak, indicated in Fig.~\ref{fig:fig1}~(b) by the arrow and the question mark.
%
Its origin still needs to be clarified, but it could be related to some specific manifestations of the internuclear coupling in crystals, not observable in traditional NMR, because under high magnetic  field these interactions are truncated by the Zeeman effect.
Another discrepancy between the data and the model concerns the zero-field spectrum obtained in the configuration where OMF is oriented perpendicular to the sample surface. This could result from the incomplete thermalization within CdTe  NSS. 

The paper is organised as follows. In the next section we present the  structure of the studied sample and the main principles of the the warm-up spectroscopy. More information about experimental protocol can be found in Appendix. The cluster model that we develop to calculate the warm-up spectra is introduced in Sec.~\ref{sec:model}. In Section \ref{sec:exp} the experimentally measured spectra are presented and compared to the model predictions. Section \ref{sec:concl} summarizes the results and points out some still unresolved issues.

\section{Sample and experimental technique}
\label{sec:sample}
 We study a $30$-nm-wide CdTe quantum well (QW) sandwiched between Cd$_{0.95}$Zn$_{0.05}$Te barriers. The top (bottom) barrier is $93$(1064)-nm-thick. The sample is grown by molecular beam epitaxy on a $[100]$ Cd$_{0.96}$Zn$_{0.04}$Te substrate. Nuclear spin dynamics in this structure has been studied in Ref.~\cite{Gribakin2024arXiv}. An optical study of a very similar QW can be found in Ref.~\cite{mikhailov2023arXiv}.   These studies confirm high quality of this type of structures and negligibly small exciton localization energy.

 %
%
%
%
%
%
  {Though the sample is nominally undoped}, low-temperature  photoluminescence (PL) measurements reported in Ref.~\cite{Gribakin2024arXiv} indicate some unintentional doping. 
%
In    wide CdTe/CdZnTe QWs excitonic states are known to be quantized  as a whole, and in the studied structure the first excited state of the free exciton emits light at $1.602$~eV  \cite{daubigne_optical_1990}. 
This resonance is chosen to monitor  PL polarization state in the detection part of the warm-up spectroscopy protocol described below.
Importantly, although we study the NSS in the quantum layer in a heterostructure, the experiments presented here do not call on the QW-specific properties. 
Therefore, the NMR spectra reported below should be relevant for bulk CdTe crystals.


The experimental method that we implement to measure NMR spectrum is an all-optical technique, termed warm-up spectroscopy. It has  been  applied previously  in GaAs\cite{Litvyak_warm-up_2021, Litvyak_local_2023}.  A typical measurement (see also Fig.~\ref{fig:protocol} and Section~\ref{sec:appendix}) consists of five steps : 

(i) Complete depolarization of the NSS by irradiation with OMF at $f=3$~kHz during $10$~s in zero static field and "in the dark".

(ii) Optical cooling of the NSS during $700$~s in the presence of the longitudinal magnetic field $B_p=150$~G (Faraday geometry).  The excitation laser emits at $1.82$~eV (far above QW barriers) and delivers $15$~mW.  It is circularly polarized and focused on a $100$~$\mu$m-diameter spot on the sample surface.

(iii) Adiabatic demagnetisation "in the dark" from $B_p$ to zero magnetic field within $20$~ms, which results in additional cooling of the NSS \cite{AbragamProctor, Vladimirova2018}.

(iv) Application of the OMF $B_\mathrm{OMF}=1$~G at frequency $f$ during $t_{\mathrm{OMF}}=1$~s. OMF field is applied either along the growth axis, or in the sample plane, either in   {presence or in absence} of the static magnetic field $B$.  Application of the OMF warms the NSS, the efficiency of this process depends on the difference between the OMF frequency $f$ and the frequency of nuclear spin resonances. 

(v) The last step  consists in the measurement and quantitative analysis of the observed increase of the NSS temperature. To do so we switch on  the transverse magnetic field $B_m=1.2$~G (Voigt geometry) and  the optical pumping. If the NSS is still cold enough, then an effective field $B_N(f)$ acting on the electrons resident in the QW will be created via hyperfine interaction. The magnitude of this field can be extracted from the PL polarisation degree as described in Appendix~\ref{sec:appendix}. 
The ratio between $B_N(f)$ and  the nuclear field $B_{N0}$ that we determine from the strictly identical  experiment but without OMF, allows us to determine the OMF absorption rate at a given frequency as:
\begin{align} \label{eq:warmup_rate}
    \frac{1}{T_{\mathrm{OMF}}}(f) = \frac{1}{t_{\mathrm{OMF}}}
    \mathrm{ln\left( \frac{B_N(f)}{B_{N0}} \right)}.
\end{align}

  {By repeating the entire  protocol for frequencies ranging from $30$~Hz to $22$~kHz we obtain the NMR spectrum at a given static field and orientation of the OMF.} The detailed description of the procedure that allows as to determine $B_N(f)$ and  $B_{N0}$ is given in Appendix~\ref{sec:appendix}. All the measurements presented throughout this work are conducted at  at $T=7$~K. 

An example of the NMR spectrum measured at $B=0$ and OMF $\parallel [001]$ is shown in Fig.~\ref{fig:fig1}~(b). The interpretation of this spectrum within the cluster model, as well as the measurements in the presence of static fields and OMF with different orientations are reported in Sec.~\ref{sec:exp}. 

\section{Absorption of the electromagnetic radiation by the NSS: the cluster model}
\label{sec:model}
In this section we develop a model that allows us to calculate the absorption of the electromagnetic radiation by the NSS in ZULF regime under the hypothesis that  this absorption is determined by nuclear clusters.
We start by defining the cluster as the set of magnetic isotopes located in a given way relative to each other. The nearest neighbors of a nucleus included in a cluster are either nuclei from the same cluster or non-magnetic nuclei. 

\begin{figure}
	\includegraphics[width=3.4in]{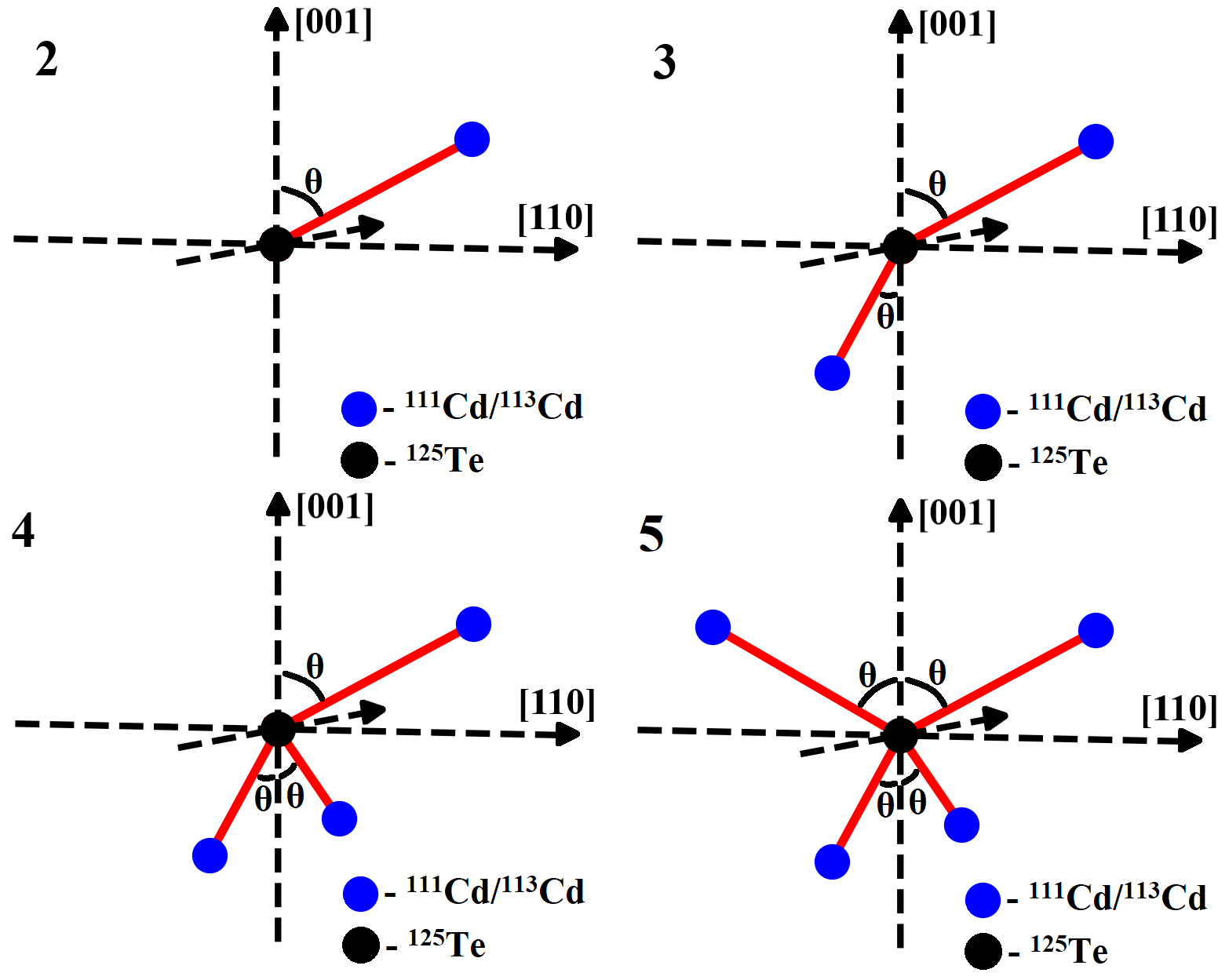} 
	\caption{ Sketch  of the most frequent nuclear magnetic clusters in CdTe, with $N=2$ to $N=5$ spins. Blue circles are Cd isotopes, black circles are $^{125}$Te, red lines indicate interatomic bonds, $\mathrm{\theta}$ is the angle between interatomic bond and crystallographic axis.
}
	\label{fig:figCluster}
\end{figure}

We rank the clusters by their size, because, as will be shown below, their abundance in the crystal drops quickly as their size increases. Moreover, the clusters containing the same number, $N$,  of magnetic nuclei may differ in the location of their nuclei relative to each other and to the crystal axes.  The absorption of the electromagnetic radiation depends on both cluster size and its configuration.  Examples of the clusters of different sizes are sketched in Fig.~\ref{fig:figCluster}. The number of possible cluster configurations, $C_N$, increases rapidly with $N$, see Table \ref{tab:tabCluster}  \footnote{Only clusters with single $^{125}$Te are considered in this work, since it has much lower abundance than magnetic isotopes of Cd.}. 
For example, a cluster containing two nuclei ($N=2$) can be formed in four different ways, because every nucleus in   {the} zinc-blend lattice of CdTe has four nearest neighbors. If one takes into account two magnetic  isotopes of Cd, then the number of the configurations doubles, $C_N=8$. 

In general,  for the cluster with $N$ nuclei   {its appearance probability (or, equivalently, abundance)}, is given by the sum  of the appearance probabilities over all possible configurations:
\begin{equation}
P_N=\sum_{l=1}^{C_N}{P_{Nc}(l)},
\label{eq:PN}
\end{equation}
where the appearance probability of the $l$-th cluster with $N$ magnetic isotopes, $P_{Nc}(l)$, is given by
\begin{equation}
{P_{Nc}(l)}=\frac{1}{N}{\prod_{i=1}^{N}{A(i)\cdot (1-P_m(i))^{4-G_N(i)}}}.
\label{eq:PNc}
\end{equation}
Here $i$ is the index that goes over all magnetic isotopes in the $l$-th cluster, $A(i)$ is the abundance of the $i$-th isotope, $P_m(i)$ is the probability of the $i$-th isotope to have a magnetic nearest neighbor (see Table~\ref{tab:tab1}), and $G_N(i)$ is the number of magnetic nearest neighbors of the $i$-th isotope in the $l$-th configuration.

Table \ref{tab:tabCluster} summarizes the probabilities calculated from Eqs.~\ref{eq:PN}-\ref{eq:PNc}. One can see that the abundance of the clusters decreases rapidly as the cluster size increases. 
In the following we assume that only isolated single nuclei and the clusters with up to $N=5$ nuclei contribute to the absorption spectrum,
  and neglect larger clusters. We also suppose that different clusters do not interact between each other.

 \begin{table}[!h]
\begin{tabular}{|>{\centering\arraybackslash}p{0.3\linewidth}|>{\centering\arraybackslash}p{0.3\linewidth}|>{\centering\arraybackslash}p{0.3\linewidth}|}
\hline
 N  & $C_N$ &   $P_N$ \\ 
\hline 
2 &  8&  0.012   \\ 
\hline 
3 &  48 &  0.004  \\
\hline 
4 &  344 & 0.0013   \\ 
 \hline 
5       &  2544   &   0.0006  \\ 
\hline 
\end{tabular}
\caption{Number of possible configurations and the appearance probability for the clusters with different size. Only clusters with $^{125}$Te in the center are considered.}
\label{tab:tabCluster}
\end{table}

Under above assumptions, the absorption of radiation by the clusters of the size $N$ is calculated as follows.  First, for each configuration $l$  from  the total of $C_N$  cluster's configurations, a Hamiltonian accounting for Zeeman interaction, as well as for direct and indirect spin-spin interactions can be written as: 
%
\begin{multline}
\hat{H}_N(l)=h \Bigl[ \sum_{i=1}^N{\gamma^i(\vec{\hat{I}}^i\vec{B}}) + \\
\sum_{i<j}^N{\delta ^{ij} \sum_{k,s=x,y,z} {\left(  J_{ks}^{ij}+ D_{ks}^{ij} \right)} \; \hat{I}_k^i \hat{I}_{s}^j } \Bigr].
\label{eq:H}
\end{multline}

%
Here $i$ is the index that goes over all nuclei in the cluster, $h$ is the Planck constant, $\gamma^i$ and $\vec{\hat{I}}^i$ are the gyromagnetic ratio and the spin operator of the $i$-th nuclei, respectively, $\vec{B}$ is the static magnetic field. $\delta ^{ij}=1$ if $i$-th and $j$-th nuclei are nearest neighbors, and zero otherwise.
$J^{ij}$  and $D^{ij}$  are the tensors of indirect  and direct  internuclear interactions, respectively.
In the coordinate system defined by the principal axes, these tensors read:
\begin{equation}
J^{ij}=
\begin{bmatrix}
    J_{\perp} & 0 & 0 \\
    0 & J_{\perp} & 0 \\
    0 & 0 & J_{\parallel}   
\end{bmatrix}
;\;
D^{ij}=
\begin{bmatrix}
    D & 0 & 0 \\
    0 & D & 0 \\
    0 & 0 & -2D 
\end{bmatrix},
%
\label{eq:JD}
\end{equation}
where $J_{\parallel}$  and  $J_{\perp}$ are the constants of indirect interaction between $i$-th and $j$-th nearest neighbor nuclei   {along the direction of the interatomic bond and perpendicular to it}, respectively.  Their values, reported by Nolle \cite{nolle_direct_1979}, are given in Table~\ref{tab:tab2}. 
The direct interaction constant $D$ is given by
\begin{equation}
{D}=\frac{\mu_0\hbar}{4\pi}\frac{\gamma^i \gamma^j}{r_0^3}.
\label{eq:D}
\end{equation}
It characterises direct dipole-dipole interaction between $i$-th and $j$-th nearest neighbor nuclei, see Table~\ref{tab:tab2}. Here $\mu_0$ is the vacuum magnetic permeability,  $r_0$ is the distance between nearest neighbors, $r_0=0.28$~nm in CdTe. 

Tensors $J^{ij}$ and $D^{ij}$ are written in the coordinate system defined by the principal axis. It is directed along the internuclear bound, which is rotated by the angle $\theta$ with respect to the $[001]$ crystallographic axis, see Fig.~\ref{fig:figCluster}. Since in our experiments static and OMF fields are applied along the crystallographic axes  we  rotate these tensors by an angle $\theta$ for further calculations.  

The Hamiltonians $\hat{H}_N(l)$ for each cluster configuration are diagonalised numerically yieldings an energy spectrum and a set of the corresponding wavefunctions. 
The OMF induces transitions between a pair of the cluster spin states if its frequency matches the energy difference between the energy levels. 
 The probability of transition $P_{mn}$  between energy levels $E_m$ and $E_n$  is given by  $P_{mn} \propto M^2_{mn}$, where $M_{mn}=\langle \Psi _m\left|H_\mathrm{OMF}\right|\Psi_n\rangle$ is the matrix element of the Hamiltonian describing Zeeman interaction between  $ \vec{B}_\mathrm{OMF} $ and  the nuclear spins:
 \begin{equation} 
 H_{OMF}=h \sum_{i=1} ^N \gamma^i \left( \vec{\hat{I}}^i \vec{B}_\mathrm{OMF}  \right).
 \label{eq:HOMF}
 \end{equation}
 The warm-up rate associated with a given transition {is proportional to its probability and to the square of its energy}
\begin{equation} 
\label{eq:wurate_theor_mn} 
\frac{1}{T_\mathrm{OMF}}\bigg|_{mn}{\sim}
{P_{mn} \left| E_m-E_n \right|^2}.
\end{equation} 
{Assuming that the NSS is fully thermalized,} the  warm-up rate  corresponding to the $l$-th configuration of the cluster containing $N$ nuclei $\frac{1}{T_\mathrm{OMF}}\bigg|_{l}$ is then obtained by the summation of the warm-up rates over all different pairs of states ($m,n$) within the cluster. 
Finally, the warm-up rate of the ensemble of the clusters containing $N$ nuclei is calculated as a sum  over all possible
configurations, taking into account their abundance:
\begin{equation} 
\label{eq:wurate_theor} 
\frac{1}{T_\mathrm{OMF}}\bigg|_{N}=\sum_{l=1}^{C_N} P_{Nc}(l) \frac{1}{T_\mathrm{OMF}}\bigg|_{l}.
\end{equation}

\begin{figure}
	\includegraphics[width=3.4in]{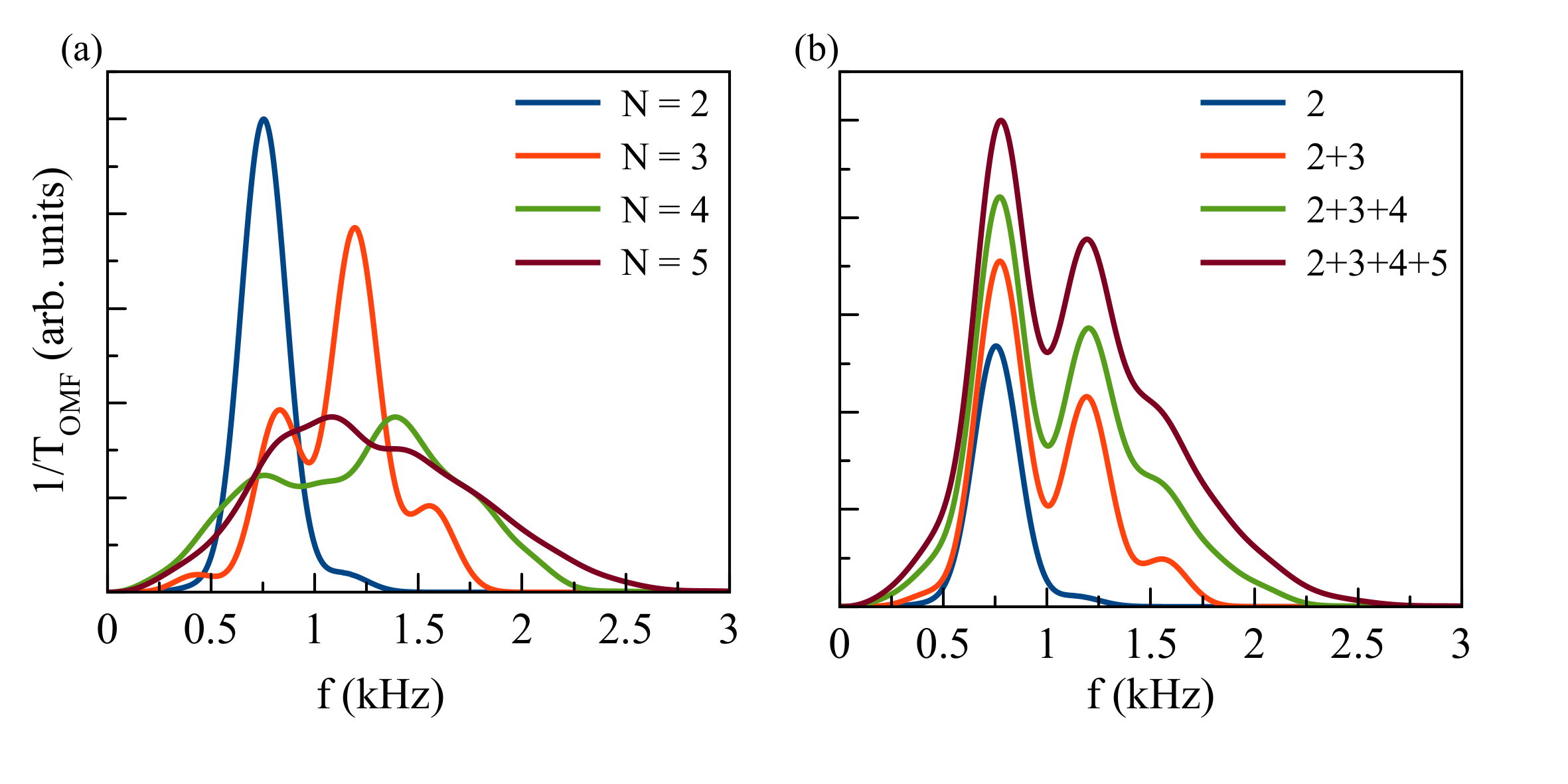}
	\caption{ Calculated absorption of the OMF energy by the nuclear spin clusters in  CdTe in zero magnetic field: (a) Absorption due to the clusters of a given size, from $N=2$ to $N=5$; 
(b) Absorption accounting for various combinations of the cluster sizes.
}
	\label{fig:figTheory}
\end{figure}

The obtained rates are transformed into spectra by convoluting them
with the Gaussians with FWHM$=0.3$~kHz.  A linewidth of this order of magnitude can be expected due to direct dipole-dipole interactions of the cluster
nuclei with other magnetic nuclei in the crystal. Since the clusters of different sizes can contribute to the total absorption of the nuclear spin system, 
we also calculate  spectra including combinations of clusters of different sizes.
The resulting zero-field warm-up spectra for CdTe NSS  are shown in Fig.~\ref{fig:figTheory}. Individual contributions of the nuclear spin clusters of size $N$ are shown in (a), while the combined absorption of the clusters with different sizes is shown in (b).

One can see that with increasing number of nuclei in the cluster the number of peaks in the spectrum grows up, the separation between them decreases, and the whole spectrum shifts towards higher energies. 
None of the spectra corresponding to the clusters of a given size matches the salient features of the measured spectrum shown in Fig.~\ref{fig:fig1}~(b), namely three distinct peaks and a high-frequency tail, but a combined contribution of all the clusters  up to $N=5$ does present a similar structure.

In the next Section we present nuclear spin absorption spectra measured for two different orientations of the OMF, in zero and low magnetic fields, and compare the data with the calculations performed within the cluster model.
\begin{figure}
	\includegraphics[width=3.4in]{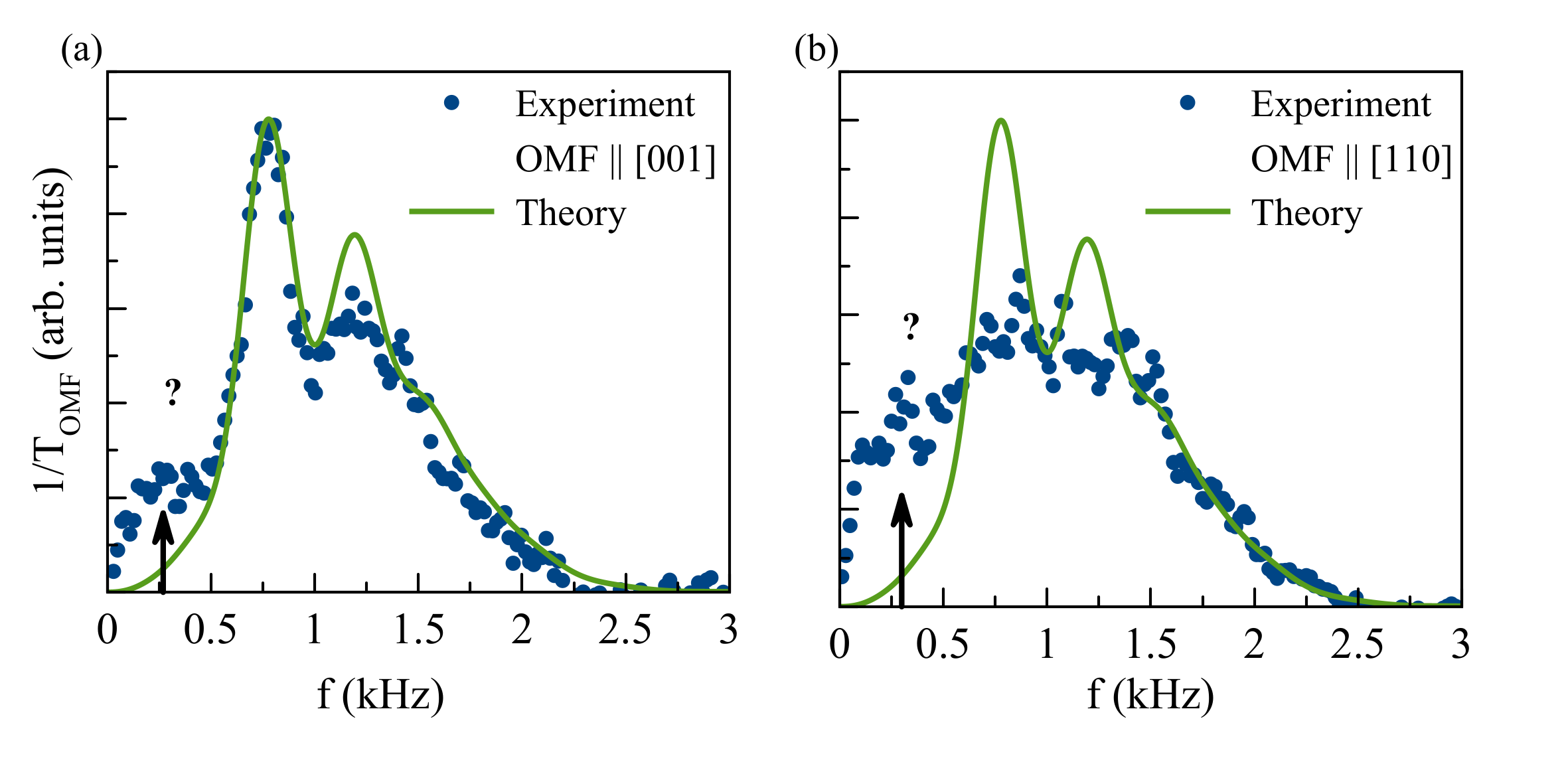}
	\caption{ Zero-field absorption spectra measured (symbols) and calculated within the cluster model assuming the contribution of all the clusters containing up to five nuclei (lines). Twe different orientation of the OMF are explored, $B_\mathrm{OMF} \parallel [001]$ (a) and $B_\mathrm{OMF} \parallel [110]$ (b).
}
	\label{fig:ExpB0}
\end{figure}

\section{Experimental results and comparison with the model}
\label{sec:exp}

Fig.~\ref{fig:ExpB0} shows the OMF absorption spectra measured in zero static field for two different orientations of the OMF, either parallel (a), or perpendicular (b) to the growth axis. The spectrum in (a) is identical to the one shown in Fig.~\ref{fig:fig1}~(b), but here the spectra are compared to the calculation (lines) within the cluster model accounting for all the clusters up to $N=5$ (see also green line in Fig.~\ref{fig:figTheory}~(b)).  
One can see that the calculated spectrum  matches relatively well the experimental one when the OMF is oriented along the growth axis (Fig.~\ref{fig:ExpB0}~(a)), except the lowest frequency peak at $\approx 0.25$~kHz. This peak is indicated by  the arrow and the question sign.
It may be related to the dipole-dipole coupling between the same isotopic species separated by large distances, because this coupling is assumed to be weak, and is not included in the model. 
Estimations including up to $5$ interacting Cd isotopes indicate that this contribution would no exceed one percent of the main absorption line spectral density.  
Thus, it may have another origin. We speculate that it could be related to some ZULF-specific manifestations of the internuclear coupling in crystals, which does not exist in traditional NMR, because under high magnetic field these interactions are truncated
by the Zeeman effect.

\begin{figure}
	\includegraphics[width=3.4in]{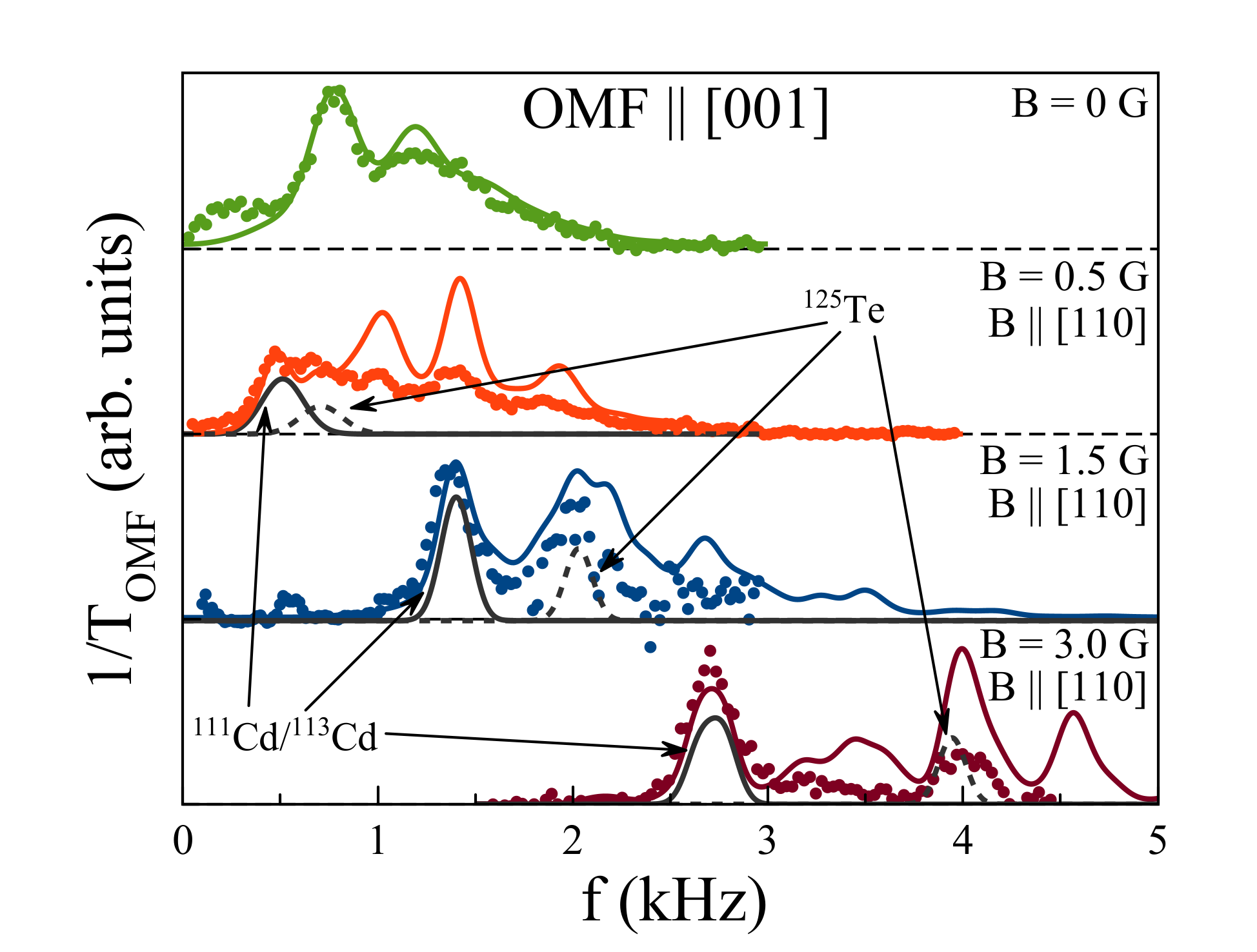}
	\caption{ Absorption spectra measured under in-plane static field ($B\parallel [110]$)  up to $3$~G (symbols). Solid lines with the corresponding color code are the spectra calculated within cluster model and accounting for clusters up to $N=5$.   {Solid and dashed gray lines represent absorption that one would expect from Zeeman splitting of individual, non-interacting Cd and Te isotopes, respectively}.
}
	\label{fig:Bperp}
\end{figure}

Remarkably, in the case where the OMF is oriented in the QW plane (Fig.~\ref{fig:ExpB0}~(b)), the spectrum has a very different shape.
While the integrated absorption is of the same order, the three-peaks fine structure is almost washed out, and only the higher-frequency tail persists.
This  result is quite surprising. Indeed, the rate of absorption at a given frequency is proportional to the imaginary part of magnetic susceptibility. The susceptibility is a second-rank tensor, which, in case of cubic symmetry, can only be a scalar. This symmetry conclusion is supported by numerical calculations within the cluster model, which gives exactly the same result for the two geometries. This significant discrepancy indicates that the model fails to describe a part of the NSS thermodynamics. We tentatively attribute this result to the incomplete thermalization within the NSS: spin clusters could retain memory of the direction in which they were initially polarized. Further research is required, however, to validate this hypothesis. 
%

Evolution of the spectrum shown in Fig.~\ref{fig:ExpB0}~(a)  upon increasing static magnetic field applied in the QW plane ($B\parallel [110]$)   up to $3$~G  is shown in Fig.~\ref{fig:Bperp}.
One can see that the zero-field absorption peaks shift towards higher frequencies and additional peaks, corresponding to non-interacting $^{111}$Cd, $^{113}$Cd and $^{125}$Te isotope's Zeeman splittings, start to form, {\it cf}  black spectra calculated for such non-interacting nuclei. 
%
%
These Zeeman contributions to the spectra are also indicated by grey arrows.
Note that at such low fields, the contributions of $^{111}$Cd and  $^{113}$Cd are still not resolved, since they have very similar gyromagnetic ratios, see Teble~\ref{tab:tab1}.
The model reproduces faithfully  the position of main absorption peaks but seems to overestimate the intensity of the blue side of the spectrum. As in the case of the $[110]$-oriented OMF absorption shown in Fig.~\ref{fig:ExpB0}~(b), this could suggest an incomplete thermalization within the NSS.

\begin{figure}
	\includegraphics[width=3.4in]{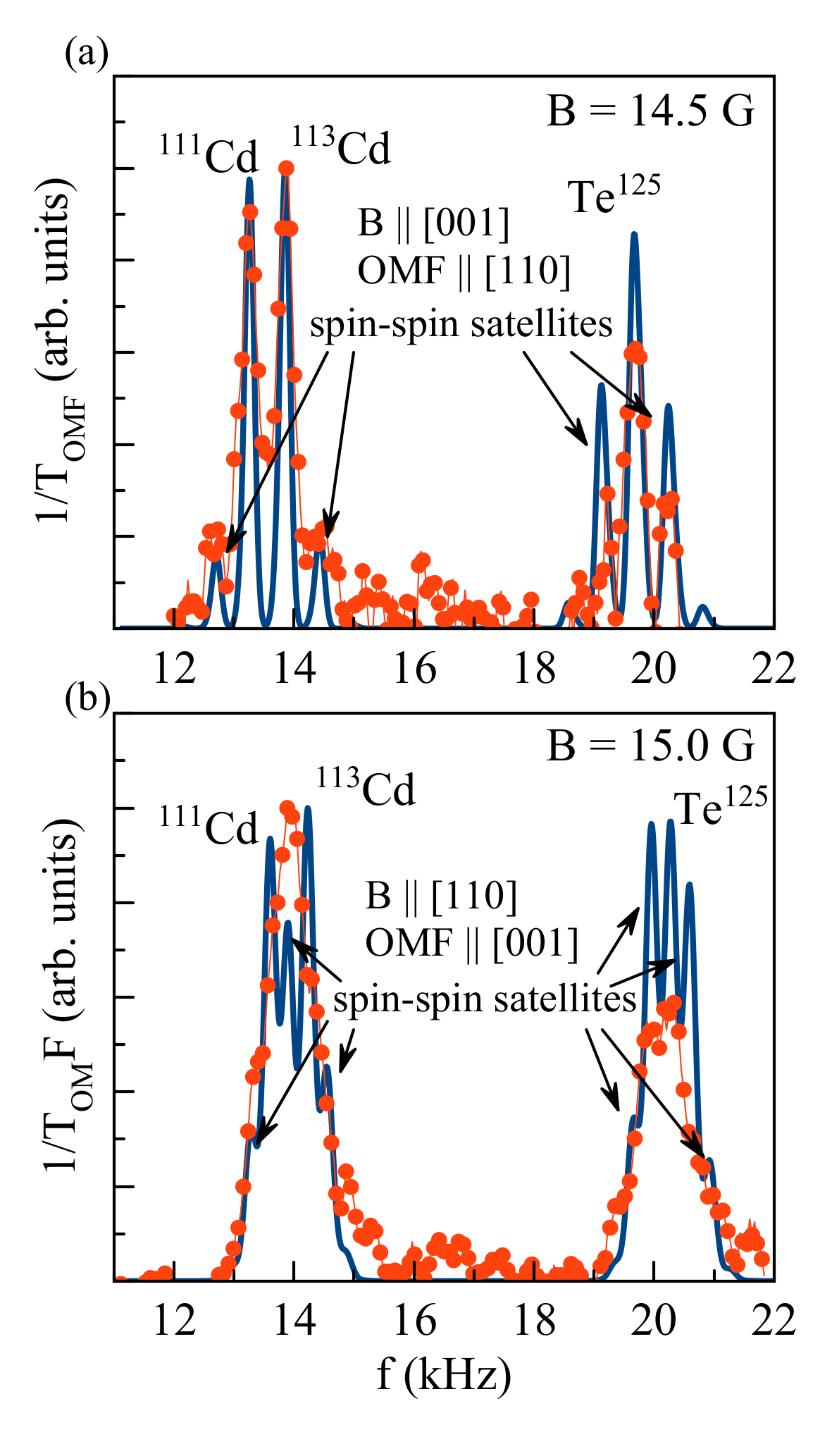}
	\caption{ Measured (symbols) and calculated (lines) absorption spectra at static magnetic fields $B\approx 15$~G for two mutually orthogonal orientations of the oscillating and static magnetic field: (a) $B= 14.5$~G and $\parallel [110]$,  OMF $\parallel [001]$; (b)  $B= 15$~G and $\parallel [001]$, OMF $\parallel [110]$.  
}
	\label{fig:B15}
\end{figure}  

At higher magnetic fields, when  Zeeman splittings of all isotopes significantly exceed internuclear interactions, a spectral structure similar to that measured by Nolle \cite{nolle_direct_1979} is expected to be observed. At $2.114$~T, Nolle observed Zeeman lines surrounded by the satellites, that he identified as being due to nuclear spin-spin interactions. The position of the satellite peaks has been shown to depend on the orientation of the static magnetic field. This experimental fact is a fingerprint of non-scalar pseudodipolar  interactions. To check the validity of this argument in our sample, it is instructive to measure magnetic field orientation-dependent nuclear spin absorption in the regime where Zeeman interaction dominates over spin-spin coupling. 

Such spectra are measured at $B\approx 15$~G for two  perpendicular orientations of the static field, see Fig.~\ref{fig:B15}. For each of these measurements the OMF orientation was adjusted to be perpendicular to the static field. One can see that  both spectra clearly demonstrate spin-spin satellites accompanying Zeeman absorption peaks. As in Ref.~\onlinecite{nolle_direct_1979}, the position of the satellites depends on the orientation of the magnetic field. This is correctly reproduced by the cluster model with $N$ limited by $5$ nuclei, and thus confirms its validity. 
We note, however, that similarly to lower-field data (Fig.~\ref{fig:Bperp}), the intensity of the high-frequency peaks corresponding to $^{125}$Te and its satellites is slightly overestimated as compared to the Cd-related low frequency part of the spectrum, { and as in zero-field experiment (Fig.~\ref{fig:ExpB0})~(b)) the fine structure is somehow washed out in the OMF$\parallel [100]$ geometry}. %
 
%

Finally, the signal associated with the very low abundance $^{123}$Te isotope  has not been identified in the measured spectra, justifying our choice to neglect it the calculations. 

\section{Conclusions}
\label{sec:concl}
We have studied, experimentally and theoretically, the absorption of electromagnetic radiation by  nuclear spins in wide CdTe/CdZnTe QW at zero and low magnetic fields.  
The reported results are not specific to heterostructures, and should apply as well to  bulk CdTe crystals.

The absorption spectra are measured using warm-up spectroscopy, a multistage technique  that comprises NSS preparation by optical pumping followed by adiabatic demagnetization,  application of the OMF which heats the NSS at a  given value of the static magnetic field, and measurement of the frequency-dependent NSS warm-up rate optically, via Hanle effect \cite{Litvyak_warm-up_2021}.

We found that CdTe NSS is characterized by much narrower absorption lines than GaAs NSS, which we have studied previously \cite{Litvyak_local_2023}. 
The main characteristics of the spectra, namely,  the fine structure observed in zero magnetic  field, as well as the satellite lines that emerge under magnetic field around  the Zeeman lines of the three main magnetic isotopes,  are understood in terms of internuclear coupling, on the basis of the "cluster model", that we developed for this purpose.
%

The model is based on the hypothesis that NSS in CdTe comprises mainly isolated noninteracting spins and small clusters consisting of up to $5$ magnetic isotopes. It accounts for direct and indirect (exchange and pseudodipolar) interactions within the clusters, while the long-range interaction between the clusters is neglected. This allows us to simplify a prohibitively complex problem resulting from the long-range character of the dipolar interaction, and reduce it to a tractable one, in order to calculate the shape of the NSS absorption spectra.

The proposed model faithfully reproduces the position of the absorption peaks in most of the studied configurations (different values of the static field, its orientation with respect to the crystal axes), while it does not have any free parameters.
%
%
The main experimentally observed features that still need to be understood include the lowest frequency peak at zero magnetic field spectrum, the overestimated intensity of the high-frequency part of the spectrum, as well as the unexpected difference between zero-field absorption spectra measured in OMF $\parallel [001]$ and OMF $\parallel [110]$ configurations.
While the two later observations are probably related to the incomplete thermalization within the dilute NSS in CdTe, the former may point on the ZULF-specific manifestations of the internuclear coupling in crystals, that can not be observed in traditional high-field NMR, where these interactions are negligibly small as compared to the Zeeman effect.

%

\section{Acknowledgements}
The authors wish to thank Denis Scalbert and Boris Gribakin for inspiring discussions. Financial support  from French National Research Agency, Grant No. ANR-21-CE30-0049 (CONUS) and from Russian Science Foundation, Grant No. 22-42-09020 is gratefully acknowledged. RA has benefited from the technical and scientific environment of the CEA-CNRS joint team "Nanophysics and Semiconductors".

\section{Appendix: Measurements of the NSS warm-up rate. }
\label{sec:appendix}
The five-stages experimental protocol described in Sec.~\ref{sec:sample} comprises the measurement stage, where nuclear field $B_N$ ~\cite{overhauser1953_Overhauser_field}, created by nuclei and acting on photocreated electrons, is extracted from the PL polarisation degree. We are interested in $B_N(t=0)$, the value of $B_N$ at $t=0$, that is right after the application of the OMF. We denote this field as $B_N(f)$ when it builds up after the application of the OMF during  $t_\mathrm{OMF}$, and as  $B_{N0}$ in the reference experiment,  were OMF is not applied. 
Once these fields are measured, the absorption rate at frequency $f$ is readily obtained via Eq.~{\ref{eq:warmup_rate}}.

\begin{figure}
	\includegraphics[width=3.4in]{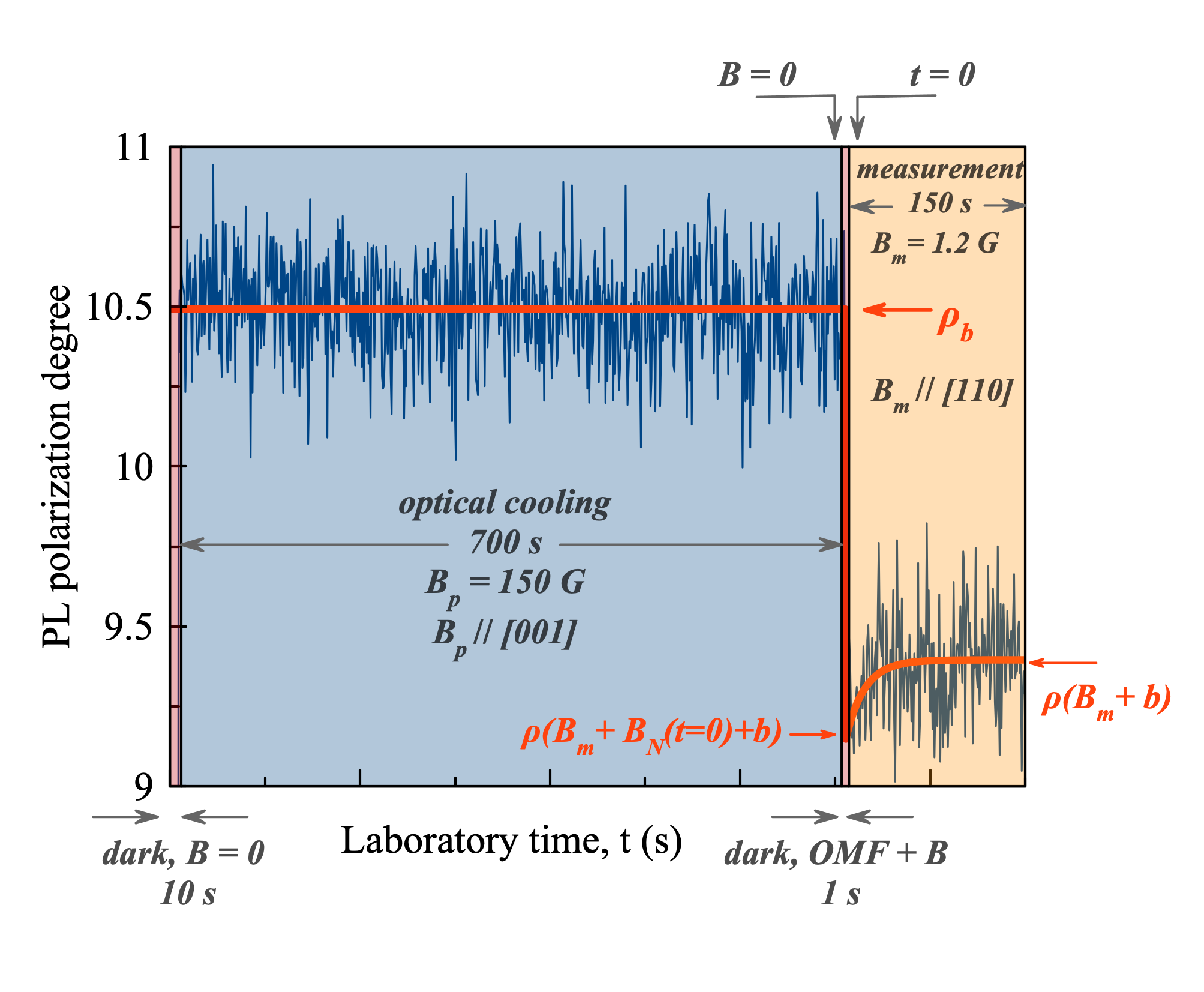}
	\caption{PL polarization degree measured (blue line) in a single warm-up experiment that allows for the determination of $B_N$ at $t=0$ by fitting (red line) Eqs.~\ref{eq:rho}~and~\ref{eq:rho2} to the data. Relevant values of the PL polarization are indicated by red arrows.
}
	\label{fig:protocol}
\end{figure} 

To determine the field $B_N(t=0)$, circularly polarized laser beam and the transverse measuring field $B_{\rm m}$ are switched on at $t=0$, and then the PL polarization degree $\rho(t)$ is recorded during $150$~s.
The polarization recorded during the entire duration of the protocol, including the pumping stage,  is illustrated in Fig.~\ref{fig:protocol}. Three important values of the PL polarization are indicated by red arrows. The polarization under optical pumping in the presence of the  longitudinal field $B_p$ is denoted as $\rho_b$, where the index indicates that it accounts for the presence of the field $b$, a  small nuclear field brought about by nuclear spin cooling in the Knight field of photocreated electrons~\cite{Litvyak_warm-up_2021}. The polarization at $t=0$ in the presence of both measurement field and the nuclear field is denoted as $\rho(B_m+B_N(t=0)+b)$. The nuclear field $B_N(t=0)$ is the one we need to extract from this measurement. Finally, the polarization $\rho(B_m+b)$ is reached at the end of the measurement stage, after complete depolarization of the NSS.  One can see that $\rho_b$ is higher than two other polarization values discussed above. This is the manifestation of the Hanle effect, that predicts depolarization of the PL in the presence of the transverse field and provides the relation between the value of the transverse field and PL polarization. 

Assuming that during measurements nuclear spin polarization decays exponentially with a characteristic time $T_1$, the PL polarization degree in the total magnetic field experienced by electrons $B_{\rm m} + B_{N}(t)+b$ can be written as \cite{OO1}:
\begin{gather}
        \rho(t) = \rho_b \frac{ B_{1/2}^{2}}{B_{1/2}^{2} \!+\! \tilde{B} ^{2}},
        \label{eq:rho}
\end{gather}
where
\begin{gather}
         \tilde{B}= \left[B_{N}(t=0) - b\right] \exp\left( - t/T_{1}\right)+B_{m} \!+\! b.
        \label{eq:rho2}
\end{gather}
Here  $B_{1/2}$ and $T_1$ are the independently measured half-width of the Hanle curve and the PL depolarization time, respectively.

Fitting  Eq.~\ref{eq:rho} to the data like those shown in Fig.~\ref{fig:protocol} allows us to determine $B_N(t=0)$.  In the case where the OMF at frequency $f$ is applied prior to measurements, we have $B_N(t=0) \equiv  B_N(f)$, and in the reference experiment without OMF we have $B_N(t=0) \equiv B_{N0}$. 
Finally, the absorption rate is obtained from the Eq.~{\ref{eq:warmup_rate}}. 

\bibliography{refs}

\end{document}